\documentclass[pra,prl,twocolumn,showpacs,floatfix,10pt,oneside]{revtex4-1}
\usepackage{graphicx}
\usepackage{dcolumn}
\usepackage{bm}
\usepackage{subfigure}
\usepackage{float}
\usepackage{multirow}
\usepackage{booktabs}
\usepackage[normalem]{ulem}
\usepackage{amsmath}
\usepackage{latexsym,amssymb}
\usepackage[mathscr]{eucal}
\usepackage[switch*]{lineno}
\usepackage{cases}
\usepackage{subeqnarray}
\usepackage{graphics}
\usepackage{lineno}
\usepackage{color}
\usepackage[bookmarks=true,
   colorlinks=true,
   linkcolor=blue,
   urlcolor=blue,
   citecolor=blue,
   bookmarks=true,
   hyperindex=true
]{hyperref}

\begin{document}

\allowdisplaybreaks[1]

\title{Impacts of generalized uncertainty principle on the black hole thermodynamics and phase transition in a cavity}

\author{Zhong-Wen Feng\textsuperscript{1}}
\altaffiliation{Email: zwfengphy@163.com}
\author{Xia Zhou\textsuperscript{1}}
\author{Shi-Qi Zhou\textsuperscript{2}}

\vskip 0.5cm
\affiliation{1 Physics and Space Science College, China West Normal University, Nanchong, 637009, China\\
2 School of Physics and Astronomy, Sun Yat-sen University, Zhuhai, 519082,  China}


\begin{abstract}
In this work, we conduct a study regarding the thermodynamic evolution and the phase transition of a black hole  in a finite spherical cavity subject to the generalized uncertainty principle. The results demonstrate that both the positive and negative generalized uncertainty principle parameters $\beta_0$ can significantly affect the thermodynamic quantities, stability, critical behavior, and phase transition of the black hole. For $\beta_0>0$,  the black hole forms a remnant with finite temperature, finite mass, and zero local heat capacity in the last stages of evolution, which can be regarded as an elementary particle. Meanwhile, it undergoes one second-order phase transition and two Hawking-Page-type phase transitions. The Gross-Perry-Yaffe phase transition occurs for both large black hole configuration and small black hole configuration. For $\beta_0<0$, the Gross-Perry-Yaffe phase transition occurs only for large black hole configuration, and the temperature and heat capacity of black hole remnant is finite, whereas its mass is zero. This indicates the remnant is metastable and would be in the Hawking-Page-type phase transition forever. Specifically, according to the viewpoint of corpuscular gravity, the remnant can be interpreted as an additional metastable tiny black hole configuration, which never appears in the original case and the positive correction case.
\end{abstract}
\maketitle
\section{Introduction}
The Planck scale is well known as the minimum scale in nature. In the vicinity of it,  many works expected that the quantum theory and gravity would merge into a theory of quantum gravity (QG) \cite{cha1,cha2,cha3,cha4}. Therefore, the Planck scale can be regarded as a demarcation line between the classical gravity and the QG. For a long time, it is believed that the properties of different physical systems would be properly changed at the Planck scale due to the effect of QG. For example, when the Heisenberg Uncertainty Principle (HUP) approaches the Planck scales, it should be modified to the so-called Generalized Uncertainty Principle (GUP) \cite{chp1,chp2,chp3}.  In this sense, Kempf, Mangano, and Mann \cite{chp1} proposed one of the most adopted GUP, as follows:
\begin{equation}
\label{eq1}
\Delta x\Delta p \ge \frac{\hbar }{2}\left[ {1 + {\beta _0}\frac{{\ell _p^2\Delta {p^2}}}{{{\hbar ^2}}}} \right],
\end{equation}
where $\ell_p$ is the Planck length, $\beta_0$ is a dimensionless GUP parameter, while that $\hbar$ corresponds to the reduced Planck constant. Meanwhile, inequality~(1) is equivalent to the modified fundamental commutation relation   $\left[ {{x_i},{p_i}} \right] = i\hbar {\delta _{ij}}\left[ {1 + {\beta _0}} \right.$ $\left. {\ell _p^2{p^2}{\hbar ^2}} \right]$ with the position operator $x_i$ and momentum operator $p_i$. For the sake of simplicity, the GUP parameter is always taken to be positive and of the order of unity, so that the results are only efficient at the Planck length. However, if this assumption is ignored, Eq.~(\ref{eq1}) can be treated by phenomenology for constraining bound of GUP parameter from experimental and observational, such as Gravitational bar detectors, Electroweak measurement, $^{87}${\rm{Rb}}  cold-atom-recoil experiment, Shapiro time delay \emph{et al}. \cite{cha5,cha6,cha7,cha8,cha9,cha10,cha11,cha12,cha13,cha13+}. With this in mind, it is believed that some new physics may appear \cite{cha14}.

Despite the GUP with a positive parameter plays an important role in many physical systems, such as gravitational theory and astrophysics \cite{cha15,cha16,cha17,cha18,cha19,cha20}, black hole physics \cite{chb20,cha21,chc1,cha22,chb22,chb23,cha23,cha24,cha25,chb25,chb26,chb27}, cosmology \cite{cha26,cha27,cha28,cha29,chb28,chb29}, quantum physics \cite{cha30,cha31,cha32,cha33}. It is still beneficial to investigate how GUP with the negative parameters  affect the classical theories \cite{cha7,cha59,cha34,cha35+}. Recently, it has been proposed that the Chandrasekhar limit fails with a positive GUP parameter and leads to the mass of white dwarfs being arbitrarily large \cite{cha36,cha37}. For solving this paradoxical situation, Ong suggests taking a negative GUP parameter, which naturally restores the Chandrasekhar limit \cite{cha35}. In this regard, to be compatible with the previous works of thermodynamics of black holes, the Hawking temperatures with both positive and negative GUP parameters have been substantively revised in Ref.~\cite{cha38}, which can be expressed as follows:
\begin{equation}
\label{eq2}
T_H^{{\rm{GUP}}} \left( {{\beta _0} > 0} \right) = \frac{{M{c^2}}}{{4\pi {\beta _0}}}\left( {1 - \sqrt {1 - \frac{{{\beta _0}\hbar c}}{{G{M^2}}}} } \right),
\end{equation}
\begin{equation}
\label{eq2+}
T_H^{{\rm{GUP}}}\left( {{\beta _0} < 0} \right) =  - \frac{{M{c^2}}}{{4\pi {\beta _0}}}\left( {1 + \sqrt {1 - \frac{{{\beta _0}\hbar c}}{{G{M^2}}}} } \right),
\end{equation}
where $\beta_0$ is the GUP parameter, $M$ is the mass of the Schwarzschild (SC) black hole, whose line element is ${\rm{d}}{s^2} =  - f\left( r \right){\rm{d}}{t^2} +  - {f^{ - 1}}\left( r \right){\rm{d}}{r^2} + {r^2}{\rm{d}}{\theta ^2} + {r^2}{\sin ^2}\theta {\rm{d}}{\varphi ^2}$ with $f\left( r \right)=1 - {{2GM} \mathord{\left/ {\vphantom {{2GM} r}} \right.
 \kern-\nulldelimiterspace} r}$. It is notable that the modified Hawking temperatures reproduce the original cases for ${\beta _0} \to 0$. Moreover, based on the first law of black hole thermodynamics  ${\rm{d}}M = T{\rm{d}}S$, the GUP corrected entropy associated with Eq.~(\ref{eq2}) and Eq.~(\ref{eq2+}) read
\begin{align}
 \label{eq3}
{S^{{\rm{GUP}}}}\left( {{\beta _0} > 0} \right) & =
  2\pi \left\{ {\frac{{G{M^2}}}{{{c^3}\hbar }}\left( {1 + \sqrt {1 - \frac{{{\beta _0}}}{{G{M^2}}}} } \right)} \right.
 \nonumber \\
&  \left. { - \frac{{{\beta _0}}}{{{c^2}}}\ln \left[ {M\left( {1 + \sqrt {1 - \frac{{{\beta _0}}}{{G{M^2}}}} } \right)} \right]} \right\},
\end{align}
\begin{align}
 \label{eq3+}
{S^{{\rm{GUP}}}}\left( {{\beta _0} < 0} \right)& =
  2\pi \left\{ {\frac{{G{M^2}}}{{{c^3}\hbar }}\left( {1 + \sqrt {1 + \frac{{{\beta _0}}}{{G{M^2}}}} } \right)} \right.
  \nonumber \\
 & \left. { - \frac{{{\beta _0}}}{{{c^2}}}\ln \left[ {M\left( {1 + \sqrt {1 + \frac{{{\beta _0}}}{{G{M^2}}}} } \right)} \right]} \right\},
\end{align}
where the logarithmic corrections on the right-hand side of Eq.~(\ref{eq3}) and Eq.~(\ref{eq3+}) are consistent with the expectation of QG theories \cite{cha35+}. The original area law of the entropy  $S = {{4\pi G{M^2}} \mathord{\left/ {\vphantom {{4\pi G{M^2}} {{c^3}\hbar }}} \right. \kern-\nulldelimiterspace} {{c^3}\hbar }}$ is recovered in HUP. According to Eq.~(\ref{eq2})-Eq.~(\ref{eq3+}), it is worth noting that whether ${\beta _0} > 0$  or ${\beta _0}< 0$, the pictures of Hawking
radiation are deviate from the classical one (see Refs.~\cite{cha35,cha38}).

On the other hand, the thermodynamic properties and evolution of black holes can be described not only by the Hawking temperature and the corresponding entropy but also by their phase structures and critical phenomena \cite{cha38+,cha38++,cha39,cht1,cht2,cha40,cha41,cha42,cha43,cha44,cha45,cha46,cha46+,chxx58,chb49,chb52,chxx59,chxx60,cha47,cha48,cha49,chb47,chb48,chb49+,chb50+,chb51+,chb52+,chb53+}. To our knowledge, the study of the thermodynamic phase transition of the black hole starts with Hawking and Page, who pointed out the existence of a thermodynamic phase transition (known as ``Hawking-Page phase transition") in the asymptotically anti-de Sitter (AdS) SC black hole when its temperature reaches a certain value. This pioneering work demonstrates the deeper-seated relation between confinement and deconfinement phase transition of the gauge field in the AdS/CFT correspondence \cite{cha39}. Moreover, this correspondence can be used to investigate the behaviour of various condensed matter phenomena \cite{cht1,cht2}. Therefore, inspired by the classical theory of Hawking-Page phase transition, the similar investigations were extended to a variety of complicated AdS spacetimes \cite{cha40,cha41,cha42,cha43,cha44,cha45,cha46,cha46+}. Beyond these achievements, people are also exploring the phase structures and critical phenomena of the non-AdS spacetimes. The biggest obstacle to achieving this is the lack of reflective surfaces (e.g., AdS term), which leads to the thermodynamic instability of non-AdS black holes. To overcome this issue, York suggests placing the non-AdS black holes inside a spherical cavity, so that the boundary of the spherical cavity can ensure the black holes are in a quasi-static thermally stable structure and makes the study of their phase behavior possible. In Ref.~\cite{chxx58}, York showed that the stable structure of the Schwarzschild black in a spherical cavity is similar to those of Schwarzschild AdS black hole. Subsequently, The phase behavior of RN black holes in the cavity is shown to have extensive similarity to that of Reissner-Nordstr\"{o}m (R-N) AdS black holes in the a grand canonical ensemble \cite{chb49}. In Refs.~\cite{chb52,chxx59,chxx60}, the phase structure and critical phenomena of a class of brane spacetimes in cavity are been investigated by the similar method. Those results the existence of Hawking-Page phase transitions in these thermodynamic systems. Moreover, by studying the thermodynamic properties of bosonic stars and hairy black holes in a cavity, it is found that they are very similar to holographic superconductors in AdS gravity \cite{chxx61,chxx62}. In additional, the author in Ref.~\cite{cha47} found a non-equilibrium second-order phase transition in the charged R-N spacetime. Subsequently, the phenomena of phase transition of the Kerr-Newman black hole was proved by Davies \cite{cha48}. By utilizing the path-integral formulation of Einstein's theory, the Gross-Perry-Yaffe (GPY) phase transition, which occurs for a hot flat space decays into the large black hole state \cite{cha49}. So far, the thermodynamic phase transition and critical phenomena of black holes are still a topic of concern \cite{chb50,chb51,chb53,chb54,chb55,cha58+,cha58++,cha59++,cha60++,cha61++,chf1,chf2}. More recently, the quantum gravity corrections to the thermodynamic phase transition and critical behavior of black holes have attracted a lot of attention \cite{cha49+,cha50,cha51,cha52,cha53,cha54,cha55}. In particular, when considering the effect of GUP, the modified thermodynamics of black holes in cavities are different from the original case, whereas the corresponding thermodynamic phase transition and critical behavior are similar to those of the AdS black hole \cite{cha56,cha57}. Hence, those results works may provide a new perspective for thermodynamic properties and evolution of black holes.

Recently, many works showed that the negative GUP parameters may appear in the nontrivial structures of spacetimes, e.g., the discreteness of space \cite{cha58,cha59x,cha60+,cha61+}. This indicates that the spacetimes with negative GUP parameters have different properties from that of positive GUP cases. As we know, the thermodynamic phase transitions and critical behavior of spacetimes are related to their structure. Therefore, it is believed that the negative GUP parameter
could lead to many new physical phenomena and results. To this end, the purpose of this paper is to explore how the positive/negative GUP effect changes the thermodynamic properties of the SC black hole. However, most investigations pertain to the positive GUP parameters, while the cases of negative GUP parameters case have seen comparatively little development. To this end, we would like to consider this issue and study the local thermodynamic evolution, critical behavior, and phase transition of SC black hole in the framework of GUP with positive/negative parameter, respectively. It turns out that, the positive/negative GUP parameter can changes the thermodynamics and phase structure of black holes in varying degrees, which are different from those of the standard cases.

The rest of this paper is organized as follows: in section~\ref{sect2}, we investigate the GUP corrected Hawking temperature and the specific heat of SC black hole in a cavity. Then, the issue of black hole remnants and the corresponding thermodynamic stability are discussed. According to the modified local thermodynamic quantities, the thermodynamic critically and phase transition of SC black hole is analyzed in detail in section~\ref{sce3}. The conclusion and discussion are contained in
section~\ref{sce4}. Throughout this paper we adopt the convention $\hbar  = c = {k_B} = 1$.

\section{The GUP corrected thermodynamic quantities in a cavity}
\label{sect2}
In order to detect the thermodynamic evolution and the phase transition of a black hole, one should enclose it in a cavity to keep it in a quasilocal thermally stable \cite{cha58}. Essentially, the boundary of the cavity acts as a reflecting surface to retain the radiation particles in this thermodynamic ensemble. Now, supposing the radius of the cavity is $R$ and using Eq.~(\ref{eq2}) and Eq.~(\ref{eq2+}), the GUP corrected local temperature of SC black hole for an observer on the cavity can be expressed as follows \cite{cha58}:
\begin{align}
\label{eq4}
&T_{{\rm{local}}}^{{\rm{GUP}}}\left( {{\beta _0} > 0} \right)  = \frac{{T_H^{{\rm{GUP}}}}}{{\sqrt {f\left( R \right)} }}
\nonumber \\
& = \frac{M}{{4\pi {\beta _0}}}{\left( {1 - \frac{{2GM}}{R}} \right)^{ - \frac{1}{2}}}\left( {1 - \sqrt {1 - \frac{{{\beta _0}}}{{G{M^2}}}} } \right),
\end{align}
\begin{align}
\label{eq4+}
& T_{{\rm{local}}}^{{\rm{GUP}}}\left( {{\beta _0} < 0} \right)  = \frac{{T_H^{{\rm{GUP}}}}}{{\sqrt {f\left( {{R}} \right)} }}
\nonumber \\
& =  - \frac{M}{{4\pi {\beta _0}}}{\left( {1 - \frac{{2GM}}{{{R}}}} \right)^{ - \frac{1}{2}}}\left( {1 - \sqrt {1 + \frac{{{\beta _0}}}{{G{M^2}}}} } \right),
\end{align}
in which the above equation  is implemented by the blue-shifted factor of the metric of the SC black hole. Mathematically, those modifications are not only sensitively dependent on mass $M$ but also the GUP parameters $\beta_0$. They respect the original local temperature  $T_{{\rm{local}}}^{{\rm{original}}} = \left( {{{{c^3}\hbar } \mathord{\left/ {\vphantom {{{c^3}\hbar } {8\pi GM}}} \right.\kern-\nulldelimiterspace} {8\pi GM}}} \right)$ ${\left( {1 - {{2GM} \mathord{\left/  {\vphantom {{2GM} R}} \right. \kern-\nulldelimiterspace} R}} \right)^{ - \frac{1}{2}}}$ in the limit $\beta_0=0$. For $\beta_0>0$, the GUP corrected local temperature $T_{\text {local}}^{\text {GUP }}$ is physical as far as the mass satisfies ${M'_{\rm{0}}} > \sqrt {{{{\beta _0}} \mathord{\left/ {\vphantom {{{\beta _0}} G}} \right. \kern-\nulldelimiterspace} G}}  = {m_p}\sqrt {{\beta _0}}$ since $T_{{\rm{local}}}^{{\rm{GUP}}} \in \mathbb{R}$. This means that, due to the effect of GUP, the black hole terminates evaporating as its mass approaches ${M'_{\rm{0}}}$, which leads to a thermodynamically inert remnant with mass $M_{{\beta _0} > 0}^{{\rm{res }}}=M_{0}^{\prime}$ and temperature is given by
\begin{equation}
\label{eq4++}
T_{{\beta _0} > 0}^{{\rm{res}}} = {T'_{\rm{0}}} = \frac{{{m_p}}}{{4\pi \sqrt {{\beta _0}} }}\left( {1 - \sqrt {1 - \frac{1}{{G m_p^2}}} } \right),
\end{equation}
where $m_p$ represents the Planck length. It should be noted that this kind of remnant is consistent with previous works \cite{chp4,chp5,chp6,chp7}. Hence, one can find that the remnant has zero specific heat (see Eq.~(\ref{eq6}) for more discussions), which means it does not exchange the energy with the surrounding space. In particular, the behaves of the remnant is more likely an elementary particle. Therefore, the temperature of the remnant can be considered as the energy of the particle \cite{cha38}. However, if $\beta_0<0$, the result shows an ``unconventional" black hole remnant, which has no rest mass but only pure temperature $T_{0}^{\prime \prime}=1/ \sqrt{4 \pi}\beta_{0}$. Despite the remnant with zero rest mass is quite different from those of previous works, it still has been discussed in Refs.~\cite{cha7,cha52,cha59x}. Meanwhile, this remnant is regarded as reasonable when considering the evolution equation of SC black hole and its sparsity of Hawking radiation. In Ref.~\cite{cha38},  the author demonstrates the black hole cannot evaporate completely in finite time, and the corresponding Hawking radiation becomes extremely sparse. In other word, at the end of evaporation, there indeed exists a metastable, long-lived remnant that approaches zero rest mass asymptotically for ${\beta _0} < 0$. Moreover, for investigating the possibility of critical behavior, it is necessary to calculate the critical points by  considers the radius of the cavity $R$ as an invariable quantity, which satisfies the following conditions
\begin{equation}
 \label{eq4+}
{\left( {\frac{{\partial {T_{{\rm{local }}}}}}{{\partial M}}} \right)_R} = 0,\quad {\left( {\frac{{{\partial ^2}{T_{{\rm{local }}}}}}{{\partial {M^2}}}} \right)_R} = 0.
\end{equation}
In above equation, we setting $R=10$ and $G=1$, the critical values of GUP parameters, the mass of the black hole, and the local temperature are
\begin{equation}
 \label{eq4++}
\left| {{\beta _{0c}}} \right|\approx 20.710, \quad M_c^{{\rm{GUP}}} \approx 4.615, \quad T_c^{{\rm{GUP}}} \approx 0.053.
\end{equation}
It is clear that the critical ratio  ${\rho _c}\left( {{\text{GUP}}} \right) = {{{M_c}\left| {{\beta _{0c}}} \right|} \mathord{\left/
 {\vphantom {{{M_c}\left| {{\beta _{0c}}} \right|} {T_c^{{\text{GUP}}}}}} \right. \kern-\nulldelimiterspace} {T_c^{{\text{GUP}}}}} \approx 1803.333$ is different from the universal ratio ${\rho _c}\left( {{\text{RN-AdS}}} \right) = {{{P_c}{\nu _c}} \mathord{\left/ {\vphantom {{{P_c}{\nu _c}} {{T_c}}}} \right. \kern-\nulldelimiterspace} {{T_c}}} = {3 \mathord{\left/ {\vphantom {3 8}} \right. \kern-\nulldelimiterspace} 8}$ for the Van der Waals fluid/RN-AdS black hole~\cite{cha40}, the ratio ${\rho _c}\left( {{\text{K-N-AdS}}} \right) = {{{P_c}{\nu _c}} \mathord{\left/
 {\vphantom {{{P_c}{\nu _c}} {{T_c}}}} \right.
 \kern-\nulldelimiterspace} {{T_c}}} = {5 \mathord{\left/
 {\vphantom {5 {12}}} \right.
 \kern-\nulldelimiterspace} {12}}$ for Kerr-Newman-AdS black hole \cite{chp8}, and the ratio ${\rho _c}\left({{\text{RainbowGravity}}} \right) = {{{M_c}{\gamma _c}} \mathord{\left/
 {\vphantom {{{M_c}{\gamma _c}} {T_c^{{\text{GUP}}}}}} \right. \kern-\nulldelimiterspace} {T_c^{{\text{GUP}}}}} \approx 266.478$ for the black hole in the rainbow gravity \cite{cha51}, where ${P_c}$, ${\nu _c}$, and ${\gamma _c}$ are the pressure, the specific volume, and the critical values of quantum gravity parameter, respectively.

\begin{figure}[h!]
\begin{center}
\includegraphics[width=10cm]{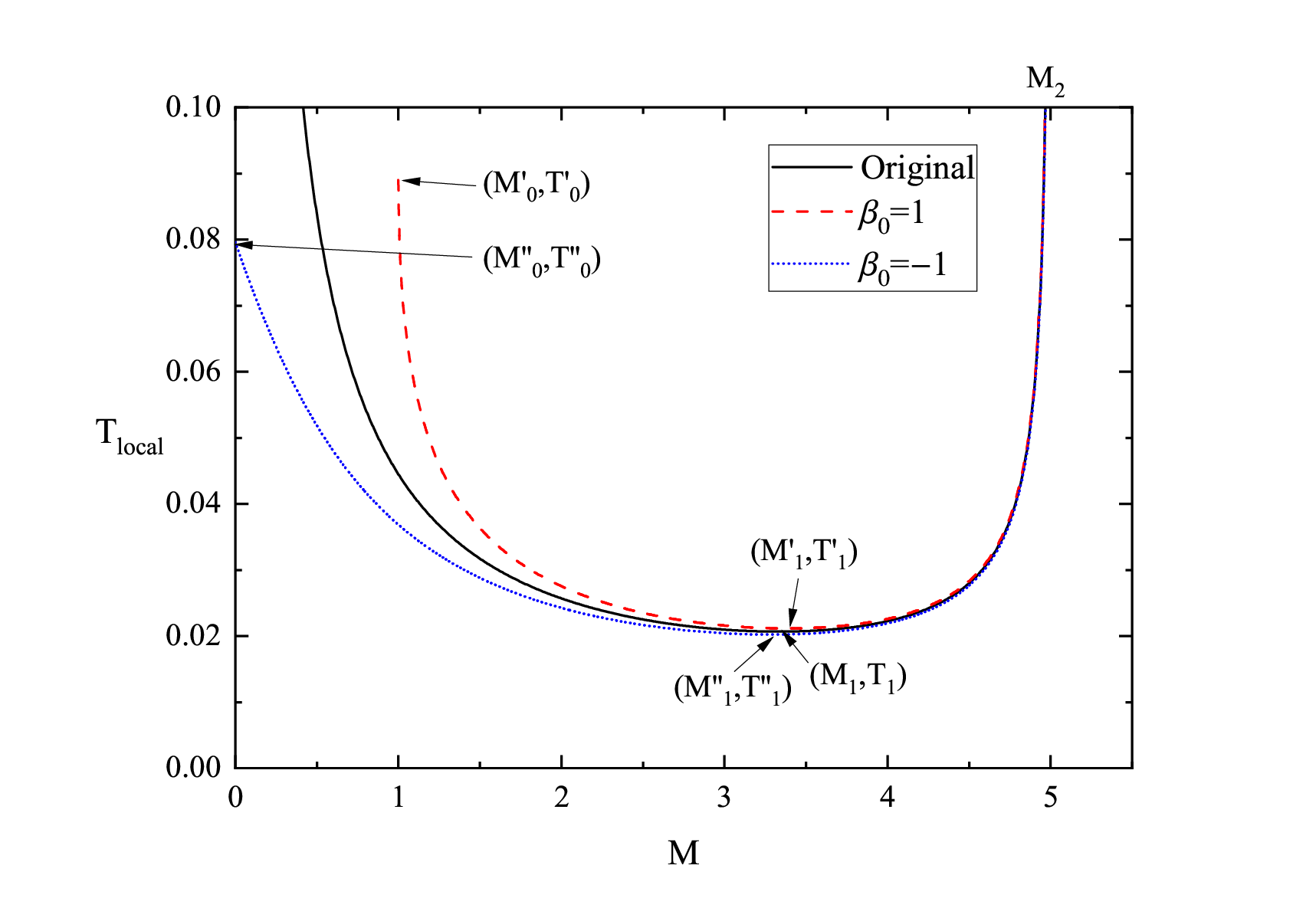}
\end{center}
\caption{The original and GUP corrected local temperature (${\beta _0} = \pm1$) as a function of mass. We set $R = 10$ and $G = 1$.}\label{fig1}
\end{figure}

Since the phase transition would occurs with  $0 < \beta_0  < \left| {{\beta _{0c}}} \right|$, hence, in order to further investigate the relationship between the local temperature and mass for different GUP parameters, e. g., $\beta_0 = \pm1$, we plot Fig.~\ref{fig1} by fixing $R=10$ and $G=1$.

As seen from Fig.~\ref{fig1}, the black solid curve corresponds to the original local temperature while the red dashed curve and the blue dotted curve represent the GUP corrected cases with ${\beta _0}=1$ and $\beta_0=-1$, respectively. It is obvious that all three kinds of local temperature have the minimum values in the ``$T_{\text {local }}-M$" plane, which can be easy numerically obtained if needed (i. e., $\left(M_{1}, T_{1}\right)=(3.333,0.021)$ for the original case;   $\left(M_{1}^{\prime}, T_{1}^{\prime}\right)=(3.384,0.0212)$ for $\beta_0=1$; $\left(M_{1}^{\prime \prime}, T_{1}^{\prime \prime}\right)=(3.284$ $,0.020)$ for $\beta_0=-1$). Those lowest points naturally divide the evolution of the black hole into two branches. The right branch represents the early stages of evolution, one can see that the original and the GUP corrected local temperatures follow the same qualitative behavior, which diverge at $M_{2}=R/2 G$, and gradually decrease with the decrease of mass before they reach the lowest points. This implies the effect of GUP is negligible at a large scale. The left branches show the final destination of the black hole: the original local temperature diverges as $M\rightarrow0$, which eventually leads to the ``information paradox". However, under the influence of GUP, SC black hole stops radiating particles and leaves the remnant at the end of evolution. For $\beta_0>0$, the red dashed curve terminates at  $M_0^\prime = M_{{\beta _0} > 0}^{{\rm{res}}} = 1$ with the $T_0^\prime  = T_{{\beta _0} > 0}^{{\rm{res }}} \approx 0.089$, whereas the remnant  with $\beta_0 < 0$ has an infinitely small value $M_0^{\prime \prime } =M_{{\beta _0} < 0}^{{\rm{res}}}$  with the temperature  $T_0^{\prime \prime } = T_{{\beta _0} < 0}^{{\rm{res}}} \approx 0.282$, as we have discussed above.

\begin{figure}[h!]
\begin{center}
\includegraphics[width=8cm]{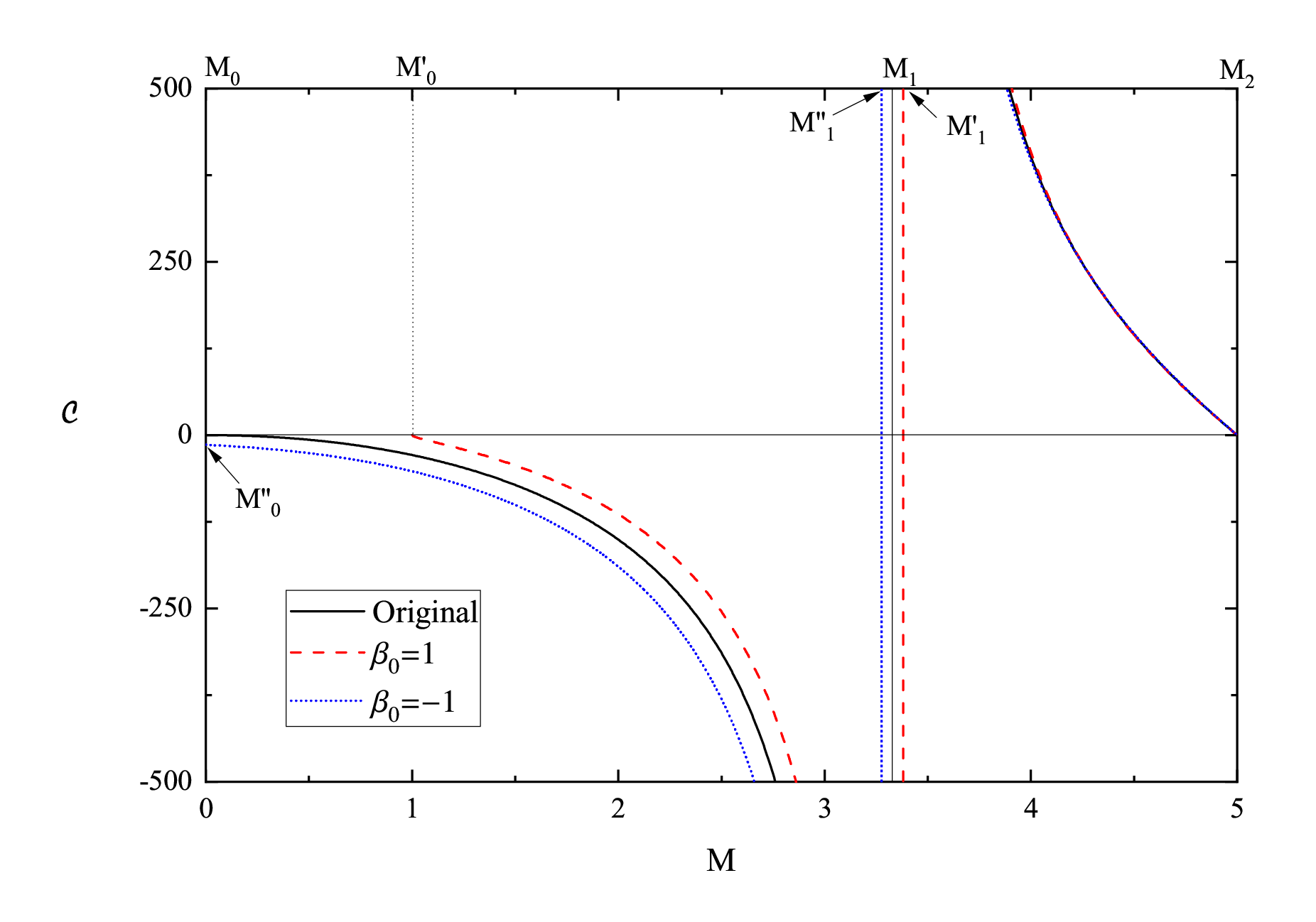}
\end{center}
\caption{The original and GUP corrected local temperature (The original specific heat and GUP corrected specific heat (${\beta _0} =\pm 1$) as a function of mass. We set $R = 10$ and $G = 1$.}\label{fig2}
\end{figure}

According to the previous works, one can classify the SC black hole as a large black hole and a small black hole depending on the two branches in Fig.~\ref{fig1}.  For confirming this viewpoint, it is necessary to study the thermodynamic stability of the black hole, which is determined by the heat capacity. Firstly, by using the first law of thermodynamics, the local thermal energy is given by
\begin{align}
\label{eq5}
E_{{\rm{local}}}^{{\rm{GUP }}}\left( {{\beta _0} > 0} \right) &= \int_{M_0^\prime }^M {T_{{\rm{local}}}^{{\rm{GUP}}}} d{S^{{\rm{GUP}}}}
 \nonumber \\
& = \frac{R}{G}\left[ {\sqrt {1 - \frac{{2{{\left( {{\beta _0}G} \right)}^{\frac{1}{2}}}}}{R}}  - \sqrt {1 - \frac{{2GM}}{R}} } \right],
\end{align}
\begin{align}
\label{eq5+}
E_{{\rm{local}}}^{{\rm{GUP }}}\left( {{\beta _0} < 0} \right) & = \int_{M_0^{\prime \prime }}^M {T_{{\rm{local}}}^{{\rm{GUP}}}} d{S^{{\rm{GUP}}}}
\nonumber \\
& = \frac{R}{G}\left( {1 - \sqrt {1 - \frac{{2GM}}{R}} } \right).
\end{align}
Note that, due to the mass of remnant of black hole, the lower limit of integration for $\beta_0>0$ is $M_0^\prime={m_p}\sqrt {{\beta _0}}$, while it becomes to $M_0^{\prime \prime } =0$ for $\beta_0<0$. When  $\beta_0\rightarrow 0$, the original local free energy  $E_{{\rm{local}}}^{{\rm{original}}} = {{R\left( {1 - \sqrt {1 - 2GM/r} } \right)} \mathord{\left/ {\vphantom {{r\left( {1 - \sqrt {1 - 2GM/r} } \right)} G}} \right. \kern-\nulldelimiterspace} G}$ is recovered. Next, according to the def\/inition $\mathcal{C} = {\left( {{{\partial {E_{{\rm{local }}}}} \mathord{\left/
 {\vphantom {{\partial {E_{{\rm{local }}}}} {\partial {T_{{\rm{local }}}}}}} \right. \kern-\nulldelimiterspace} {\partial {T_{{\rm{local }}}}}}} \right)_r}$, the GUP corrected local heat capacity within the boundary $r$ is given by
\begin{equation}
\label{eq6}
{\cal C}_{{\rm{local}}}^{{\rm{GUP }}}\left( {{\beta _0} > 0} \right) = \frac{{4\pi M {{\beta _0}} \left( {R - 2GM} \right)\sqrt {1 - \frac{{{\beta _0}}}{{G{M^2}}}} }}{{{\beta _0} - M\left( {R - GM} \right)\left( {1 - \sqrt {1 - \frac{{{\beta _0}}}{{G{M^2}}}} } \right)}},
\end{equation}
\begin{equation}
 \label{eq6+}
{\cal C}_{{\rm{local}}}^{{\rm{GUP }}}\left( {{\beta _0} < 0} \right) = \frac{{4\pi M{\beta _0}\left( {R - 2GM} \right)\sqrt {1 + \frac{{{\beta _0}}}{{G{M^2}}}} }}{{{\beta _0} + M\left( {R - GM} \right)\left( {1 - \sqrt {1 + \frac{{{\beta _0}}}{{G{M^2}}}} } \right)}}.
\end{equation}
By setting $R = 10$ and $G = 1$, the original specific heat and the GUP corrected specific heat as a function of mass for different values of $\beta_0$ is reflected in Fig.~\ref{fig2}.

As seen from Fig.~\ref{fig2}, the black solid curve for the original heat capacity ${{\cal C}^{{\rm{original}}}} = 8\pi G{M^2}\left( {R} \right.$ ${{\left. {-2GM} \right)} \mathord{\left/ {\vphantom {{\left. {2GM} \right)} {\left( {3GM - R} \right)}}} \right. \kern-\nulldelimiterspace} {\left( {3GM - R} \right)}}$ goes to zero when $M= 0$, whereas the red dotted curve for $\mathcal{C}_{{\beta _0} > 0}^{{\rm{GUP}}}$ vanishes at $M_{0}^{\prime}$, resulting in a thermodynamically inert remnant. We are more concerned about what happens when $\beta_0<0$. When the mass of the black hole is large enough, the blue dotted curve for the modified heat capacity with negative GUP parameter is large than zero, its behavior is the same as those of $\beta_0=0$ and $\beta_0>0$. However, the modified heat capacity approaches a non-zero value as the mass becomes $M_{0}^{\prime \prime}$, which is caused by the thermal interaction between metastable remnant and the environment \cite{cha59}. In particular, the similar results can be found in the framework of rainbow gravity (RG) \cite{cha50,cha60}, which implies GUP and RG have a deeper connection. Furthermore, all the curves have vertical asymptotes at the locations (i. e., $M_1$, $M_{1}^{\prime}$ and $M_{1}^{\prime \prime}$) where the local temperatures reach the minimum values. Therefore, it implies a second-order phase transition from positive to negative heat capacity at around the vertical asymptote.

Now, armed with the discussions on the modified local temperature and heat capacity, one can classify the SC black hole  into two configurations depending on its mass scale, namely, the large black hole (LBH) configuration  and the small black hole (SBH) configuration. The stability, the region of the heat capacity, and the region of the mass of small/large black holes with the positive and negative GUP parameters are depicted in Table~\ref{tab2}.
\begin{table*}[htbp]
\centering
\caption{\label{tab2} Stability and region of the mass of the small/large black hole with different values of GUP parameter.}
\begin{tabular}{c  c c  c  c}
\hline
{GUP parameter}                                 & {Branches}         & {Stability}        &{Region of heat capacity}      & {Region of mass} \\
\hline
$\beta_0 >0$  & small              & unstable         &$\mathcal{C}<0$                  & $M_{0}^{\prime}<M<M_{1}^{\prime}$\\
                                                               & large                 & stable               &$\mathcal{C}>0$                  & $M_{1}^{\prime}<M<M_{2}$\\
$\beta_0 <0$& small               & unstable          &$\mathcal{C}<0$                   &$M_{0}^{\prime \prime}<M<M_{1}^{\prime \prime}$\\
                                                              & large                 & stable               &$\mathcal{C}>0$                    & $M_{1}^{\prime \prime}<M<M_{2}$ \\
\hline
\end{tabular}
\end{table*}

From Table~\ref{tab2}, it is found that whether $\beta_0>0$ or $\beta_0<0$, the system always has one SBH with positive heat capacity and one LBH with negative heat capacity. Obviously, the stability determines that SBH cannot exist for a long time, it would decay into the remnant or the stable LBH. In this process, some interesting thermodynamic phase transitions that never appears in the original case can be found by analyzing the Helmholtz free energy of LBH and SBH.

\section{Helmholtz free energy and phase transition}
\label{sce3}
\begin{figure}[h!]
\begin{center}
\includegraphics[width=8cm]{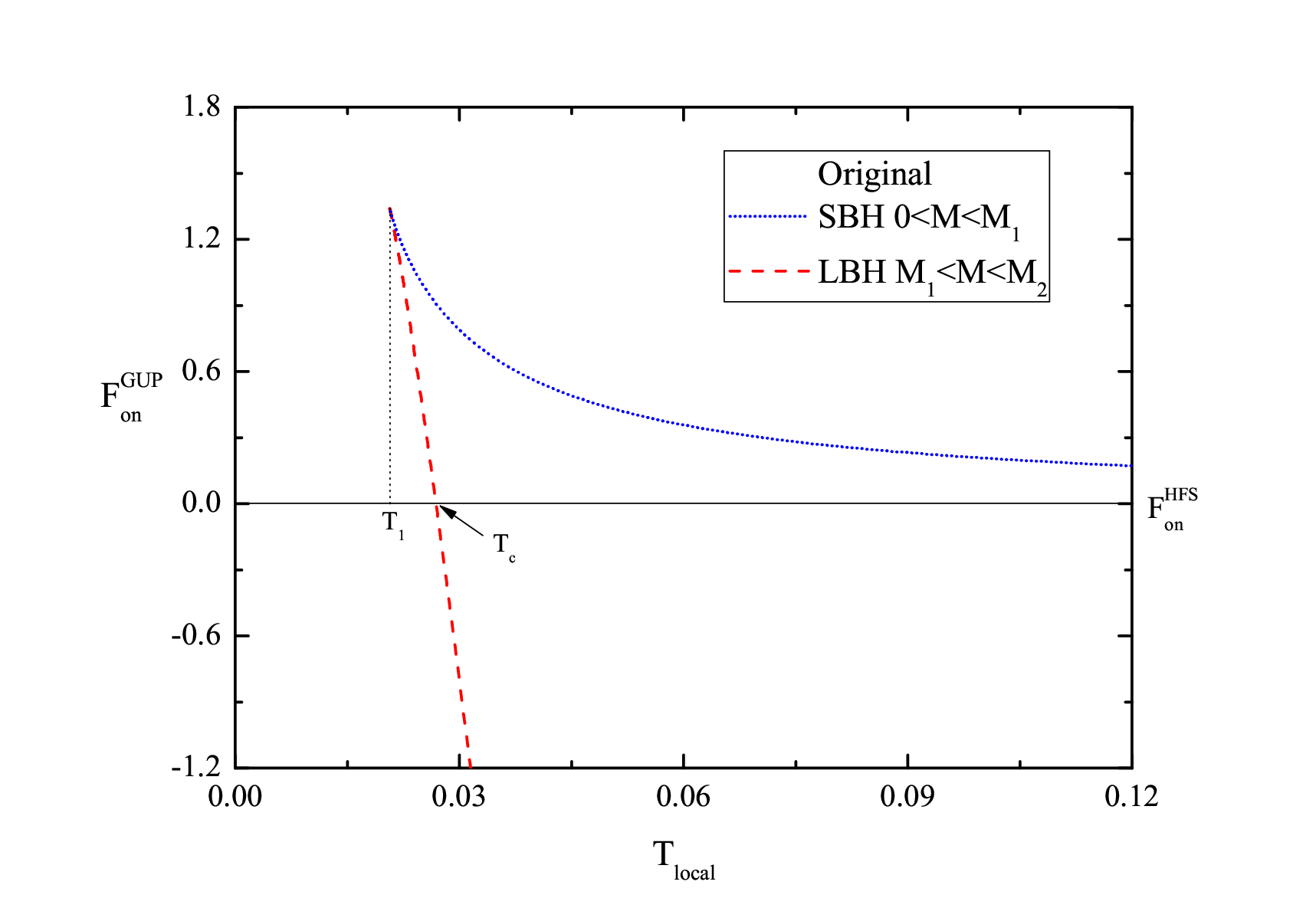}
\end{center}
\caption{The original Helmholtz free energy versus the original local temperature, We take $R = 10$ and $G = 1$.}\label{3}
\end{figure}

\begin{figure}[h!]
\begin{center}
\includegraphics[width=8cm]{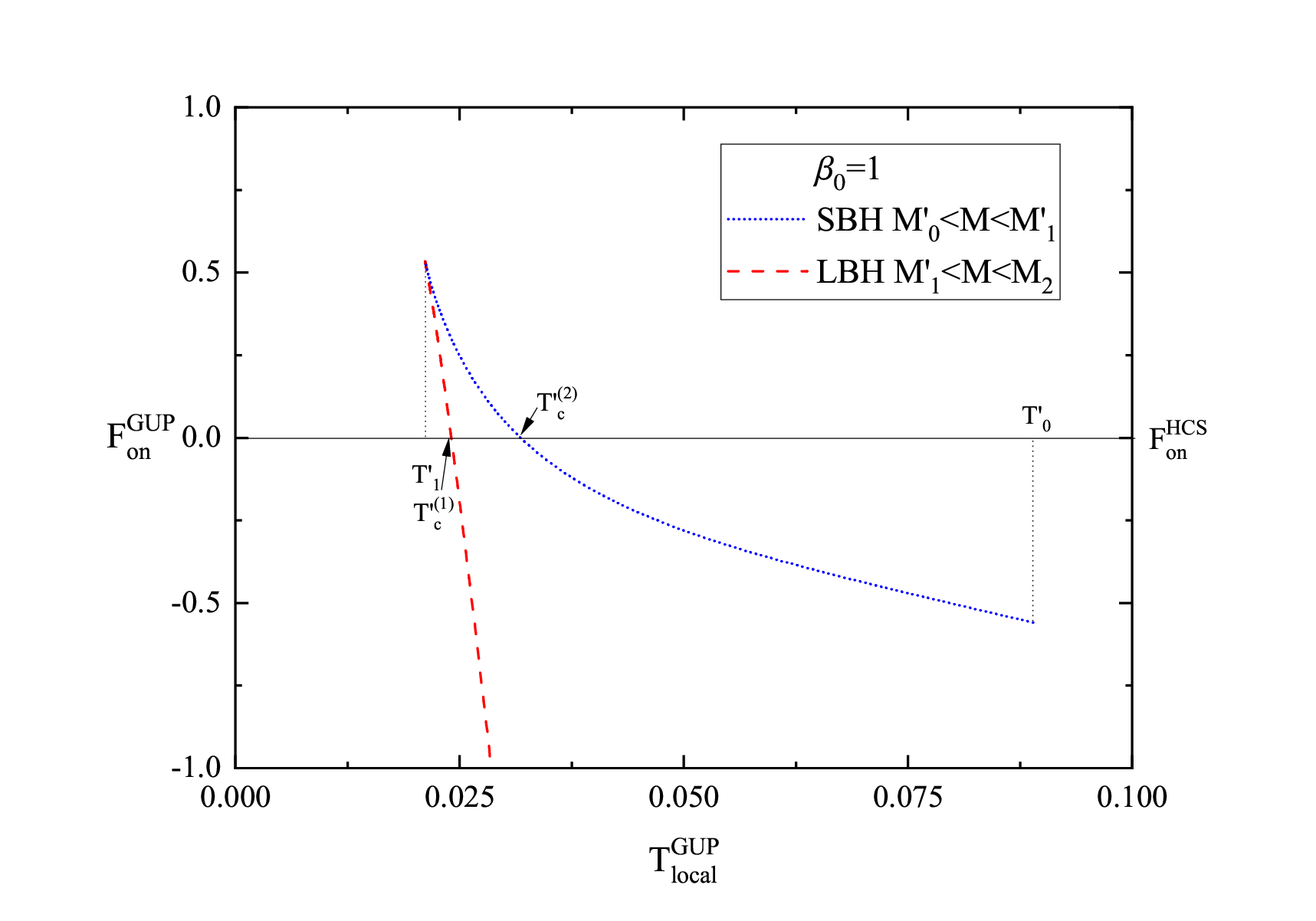}
\end{center}
\caption{The
GUP corrected Helmholtz free energy versus the modified temperature for $\beta_0=1$, We take $R = 10$ and $G = 1$.}\label{4}
\end{figure}

\begin{figure}[h!]
\begin{center}
\includegraphics[width=8cm]{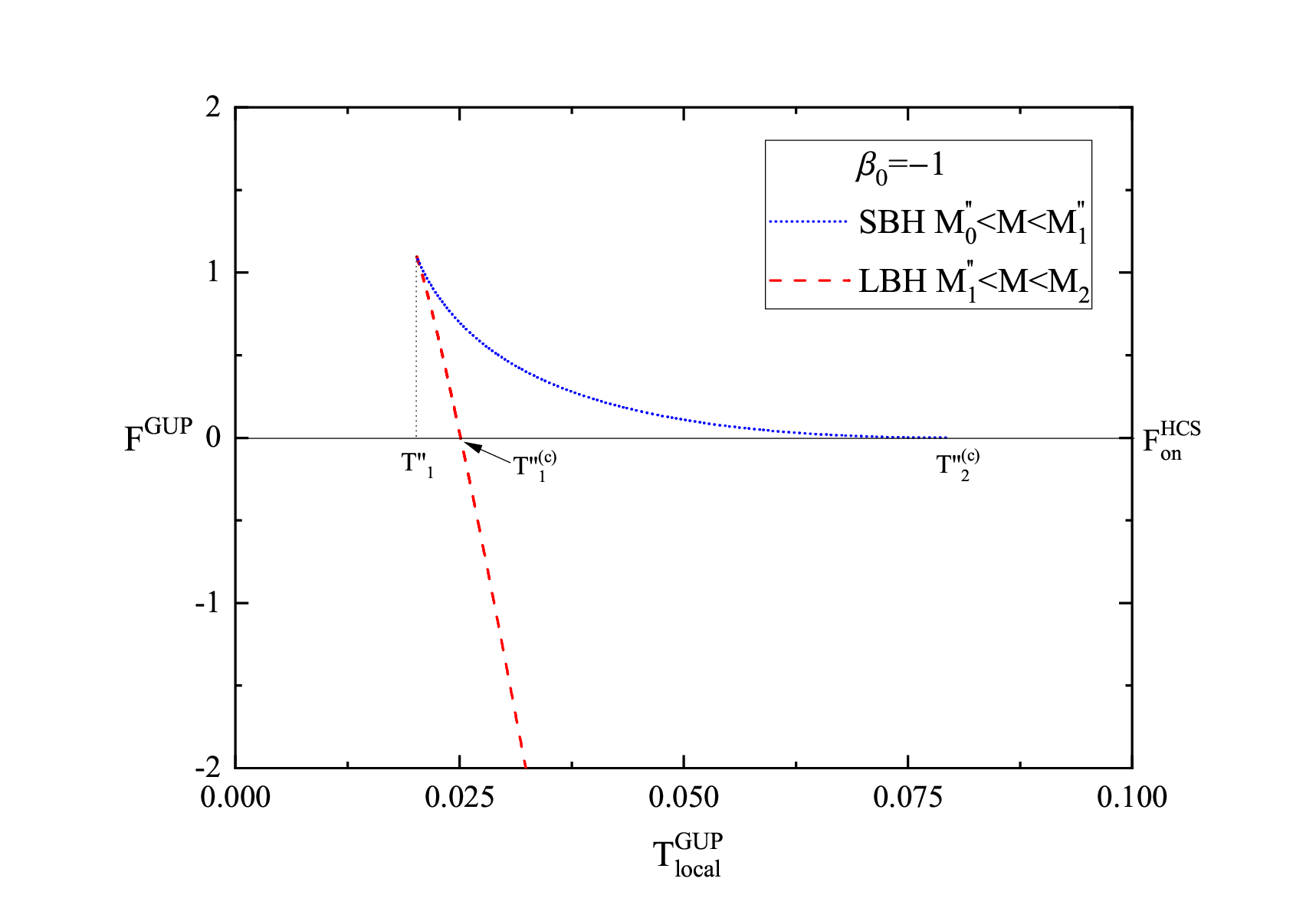}
\end{center}
\caption{The
GUP corrected Helmholtz free energy versus the modified temperature for $\beta_0=-1$, We take $R = 10$ and $G = 1$.}\label{5}
\end{figure}

In this section, it is necessary to analyze the corrections to thermodynamic criticality and phase transition due to the GUP. To this aim, one needs to calculate the Helmholtz free energy in an isothermal cavity defined as  ${F_{{\rm{on}}}} = {E_{{\rm{local}}}} - {T_{{\rm{local}}}}S$. According to Eq.~(\ref{eq3})-Eq.~(\ref{eq5+}), the GUP corrected Helmholtz free energy is given by:
\begin{align}
 \label{eq7}
F_{{\rm{on}}}^{{\rm{GUP}}}\left( {{\beta _0} > 0} \right) &= \frac{R}{G}\left( {\sqrt {1 - \frac{{2{{\left( {{\beta _0}G} \right)}^{\frac{1}{2}}}}}{R}}  - \sqrt {1 - \frac{{2GM}}{R}} } \right)
\nonumber \\
& - \frac{{M\Theta }}{{2\sqrt {1 - \frac{{2GM}}{R}} }},
\end{align}
\begin{align}
 \label{eq7+}
F_{{\rm{on}}}^{{\rm{GUP}}}\left( {{\beta _0} < 0} \right)=\frac{R}{G}\left( {1 - \sqrt {1 - \frac{{2GM}}{R}} } \right)- \frac{{M\Xi }}{{2\sqrt {1 - \frac{{2GM}}{R}} }},
\end{align}
where  $\Theta  = 1 - \left( {1 - \sqrt {1 - \frac{{{\beta _0}}}{{G{M^2}}}} } \right)\ln \left[ {M\left( {1 + \sqrt {1 - \frac{{{\beta _0}}}{{G{M^2}}}} } \right)} \right]$ and $\Xi  = 1 - \left( {1 - \sqrt {1 + \frac{{{\beta _0}}}{{G{M^2}}}} } \right)$ $\ln \left[ {M\left( {1 + \sqrt {1 + \frac{{{\beta _0}}}{{G{M^2}}}} } \right)} \right]$. In the limit  $\beta_0\rightarrow 0$, this agrees with the original Helmholtz free energy ${F_{{\rm{on}}}} = {{R\left( {1 - \sqrt {1 - 2GM/R} } \right)} \mathord{\left/ {\vphantom {{R\left( {1 - \sqrt {1 - 2GM/R} } \right)} G}} \right. \kern-\nulldelimiterspace} G}$ $ - \left( {{M \mathord{\left/ {\vphantom {M {2\sqrt {1 - {{2GM} \mathord{\left/ {\vphantom {{2GM} R}} \right. \kern-\nulldelimiterspace} R}} }}} \right. \kern-\nulldelimiterspace} {2\sqrt {1 - {{2GM} \mathord{\left/ {\vphantom {{2GM} R}} \right. \kern-\nulldelimiterspace} R}} }}} \right)$. By fixing $R = 10$ and $G = 1$, the original and modified Helmholtz free energy versus their local temperatures are presented in Figs.~\ref{3}-\ref{5}.

Fig.~\ref{3} reveals the relationship between the original local free energy and its local temperature. Note that the curve of free energy is not continuous, there is a cusp between the free energies of SBH and LBH, resulting in a second-order phase transition appears at the critical point $T_1$ corresponding to the mass  $M_1$. The horizontal line refers to the Helmholtz free energy of hot flat space (HFS)  $F_{{\rm{on}}}^{{\rm{HFS}}} = 0$ in the Minkowski spacetime without a black hole. Following the viewpoint of York, the SC black hole in an isothermal cavity should reach the thermal equilibrium between the HFS and the  black hole through the Hawking-Page phase transition \cite{cha58}. Hence, one can see a first-order Hawking-Page phase transition at  $T_c$ since the Helmholtz free energy of LBH (red dashed curve) intersects the $F_{{\rm{on}}}^{{\rm{HFS}}}$ there. On the contrary, the Helmholtz free energy of SBH tends to but never reach the line of  HFS, so the there is only one Hawking-Page  phase transition in the $F_{{\rm{on}}}^{{\rm{GUP}}}-{T_{{\rm{local}}}}$ plane. In the range of ${T_1} < T < {T_c}$, the HFS is most probable since  $F_{{\rm{on}}}^{{\rm{LBH}}}$ and $F_{{\rm{on}}}^{{\rm{SBH}}}$  are higher than  $F_{{\rm{on}}}^{{\rm{HFS}}}$. However, the relation of Helmholtz free energy changes to  $F_{{\rm{on}}}^{{\rm{LBH}}} < F_{{\rm{on}}}^{{\rm{HFS}}} < F_{{\rm{on}}}^{{\rm{SBH}}}$ for  $T > {T_c}$, which implies that the HFS and the unstable SBH eventually collapses into the stable  LBH. Moreover, according to the viewpoints in Ref.~\cite{cha49}, it f\/inds that a GPY phase transition occurs for the LBH.

Next, let us focus on Fig.~\ref{4} and Fig.~\ref{5} for the modified cases with different  GUP parameters, e. g., $\beta_0=\pm 1$. It is worth noting that spacetime is always curved due to the remnant of the black hole. Hence, there is no HFS in the framework of GUP. In the following discussion, we should use the hot curved (HCS) space to replace the HFS. More specifically, one can find that the free energy of HCS $F_{{\rm{on}}}^{{\rm{HCS}}}$ tends to zero when the mass of $M = M_{\rm res}$. More specifically,  Hence, the HFS should be replaced by the hot curved space (HCS) in the following discussion. As the counterpart of the original one, the HCS and $F_{{\rm{on}}}^{{\rm{HCS}}}$ effectively influence the phase transition of the SC black hole.

As seen from Fig.~\ref{4}, when the temperature is lower than  $T_c^{\prime \left(2 \right)}$, the behavior of the modified Helmholtz free energy is analogous to that in the case of Fig.~\ref{3}, namely, a second-order phase transition point and a Hawking-Page-type phase transition occur at the inflection point ${T^{\prime}_1}$  and $T_c^{{\prime} \left( 1 \right)}$, respectively. The Helmholtz free energy of LBH is smaller than those of SBH and HCS in the range of $T_c^{{\prime} \left( 1 \right)}< T < T_c^{{\prime} \left( 2 \right)}$, resulting in a GPY phase transition appears for the LBH. However, for $T > T_c^{{\prime} \left( 2 \right)}$, due to the effect of GUP with the positive parameter, one can explore some interesting results that different from the original one:

\begin{enumerate}
\item An additional Hawking-Page-type phase transition can be found at  $T_c^{{\prime}\left( 2 \right)}$ since the Helmholtz free energy of SBH intersect with $F_{{\rm{on}}}^{{\rm{HCS}}}$.

\item As long as $T_c^{{\prime} \left( 2 \right)} < T < {T_0^{\prime}}$, the energy of SBH decreases below $F_{{\rm{on}}}^{{\rm{HCS}}}$, which indicates that the HCS decay into SBH via the GPY phase transition. Therefore, when considering the case $\beta_0>0$, the GPY phase transition also occurs for SBH.

\item The $F_{{\rm{on}}}^{{\rm{SHB}}}$ is always higher than the Helmholtz free energy of LBH and the black hole remnant. In this case, the unstable SBH would not only decays into the stable LBH, but also into the remnant.
\end{enumerate}

Regarding Fig.~\ref{5}, one can observe that:
\begin{enumerate}

\item There is a  Hawking-Page-type phase transition from $F_{{\rm{on}}}^{{\rm{LBH}}} > 0$  to  $F_{{\rm{on}}}^{{\rm{LBH}}} < 0$ at the inflection point $T_c^{{\prime}{\prime}\left( 1 \right)}$. Besides, one can see a second-order phase transition since far the left end of the blue curve for $F_{{\rm{on}}}^{{\rm{SBH}}}$  meets the red dashed curve at at the cusp temperature $T^{{\prime}{\prime}}_1$.

\item The  $F_{{\rm{on}}}^{{\rm{SBH}}}$ can not reach the intersection point $T_c^{{{\prime}{\prime}}\left( 2 \right)}$  in a finite time. According to the arguments about the remnant without rest mass in section~\ref{sce3}, it turns out the black hole remnant caused by the negative GUP parameter is metastable and can be trapped in the Hawking-Page-type phase transition for a long time, which is consistent with the analysis of the Hawking temperature in Eq.~(\ref{eq4+}) and the specific heat in Eq.~(\ref{eq6+}).

\item  Even more remarkably, above results are reminiscent of viewpoints of the black hole in corpuscular gravity (CG), which states that the black holes can be interpreted as a condensate  at the critical point of a quantum phase transition \cite{chb1,chb2}. Therefore,  along the line of CG theory, the remnant can be interpreted as an additional metastable tiny black hole (TBH) configuration of the system. With this, both the unstable SBH and metastable TBH (or the remnant) would collapse into stable LBH eventually since the relation of  free energies obey $F_{{\rm{on}}}^{{\rm{LBH}}}<F_{{\rm{on}}}^{{\rm{TBH}}}<F_{{\rm{on}}}^{{\rm{SBH}}}$.

\item  The Helmholtz free energy of LBH is always lower than those of SBH and HCS for  $T > T_c^{{\prime}{\prime}\left( 1 \right)}$, showing the GPY phase transition only appears for the LBH.
\end{enumerate}

\section{Conclusion}
\label{sce4}

In this paper, we have explored how the GUP with positive/negative parameter affects the local thermodynamic quantities, thermal stability, and phase transitions of SC black hole in a cavity. Our results show that the positive/negative corrections have their own unique properties and are unambiguously distinguished from the original case. For $\beta_0>0$, the SC black hole leaves a thermodynamically inert remnant with a finite temperature, a finite mass, and a zero local heat capacity. By analyzing the $F_{{\rm{on}}}-T_{{\rm{local}}}$, it was  found two Hawking-Page-type phase transitions and one second-order phase transition, whereas the original case only has one Hawking-Page phase transition. As long as  $T_c^{(1)} < T < T_c^{{\prime}\left( 2 \right)}$, the free energy obeys  $F_{{\rm{on}}}^{{\rm{LBH}}} < F_{{\rm{on}}}^{{\rm{SBH}}} < F_{{\rm{on}}}^{{\rm{HFS}}}$, which implies that the GPY phase transition does not only occur for the LBH but also for the SBH. Meanwhile, the relation of free energy also shows that unstable SBH collapses into the stable  LBH or remnant eventually. However,  for $\beta_0<0$, the remnant becomes metastable, which has a non-zero  heat capacity and a finite temperature but a zero rest mass. Furthermore, this thermodynamic ensemble exists one Hawking-Page-type critical point and one second-order phase transition critical point in the range of  $T > {T_0^{{\prime}{\prime}}}$ (see Fig.~\ref{5}. Due to the $F_{{\rm{on}}}^{{\rm{SBH}}}$ is infinitely close to the horizontal line at $T = {T_0^{{\prime}{\prime}}}$, we confirmed that the remnant is metastable and be trapped in the Hawking-Page-type phase transition for a long time. Interestingly, from the CG point of view, the remnant can be interpreted as an additional TBH configuration, which never appears in the original case and the positive correction case. Lastly, we found the unstable SBH and the metastable remnant eventually collapse into stable the LBH since the free energy of LBH is always higher than those of the SBH and the remnant, so that the GPY phase transition only occurs for LBH.

For a long time, people are focus on the GUP with positive parameter. However, our work shows that the GUP with negative parameter is as important as the positive one since it can significantly affect the thermodynamics, stability, and phase structures of a black hole. These results can reasonably and consistently describe the thermodynamic evolution of a black hole, and avoid the information paradox. Specifically, the zero mass remnant can be regarded as a candidate of dark matter (see Refs. \cite{cha38,chd1}), which could be found in the further astronomical observation. Therefore, it would be very interesting to explore these phenomena in the context of black holes with the negative GUP. Finally, we only focused on how the GUP with both positive and negative parameters affect the thermodynamic properties and phase transition of SC black hole in this present work. It should be noted that our work can be applied to more generic black holes, such as SC-AdS black holes. The relevant issues will be discussed in detail in our future work.



\begin{thebibliography}{99}
\bibitem{cha1}
K. Konishia,  G. Paf\/futib, P.  Provero, Phys. Lett. B 1990,  234: 276.  \href{http://dx.doi.org/10.1016/0370-2693(90)91927-4} {DOI: 10.1016/0370-2693(90)91927-4}

\bibitem{cha2}
M. Maggiore, Phys. Lett. B 1993, 319: 83.  \href{https://arxiv.org/abs/gr-qc/9403008} {arXiv:gr-qc/9403008}

\bibitem{cha3}
L. J. Garay,  Int. J. Mod. Phys. A 1995, 10: 145. \href{https://arxiv.org/abs/hep-th/9904025} {arXiv:hep-th/9904025}

\bibitem{cha4}
G. Amelino-Camelia,  Int. J. Mod. Phys. D 2002, 11: 35.  \href{https://arxiv.org/abs/hep-th/9604045} {arXiv:hep-th/9604045}

\bibitem{chp1}
A. Kempf, G. Mangano, R. B. Mann,  Phys. Rev. D 1995, 52: 1108.  \href{https://arxiv.org/abs/hep-th/9412167} {arXiv: hep-th/9412167}

\bibitem{chp2}
 F. Scardigli,  Phys. Lett. B 1999, 452:  39.   \href{https://arxiv.org/abs/hep-th/9904025} {arXiv: hep-th/9904025}

\bibitem{chp3}
R. J. Adler,  D. I. Santiago,  Mod. Phys. Lett. A  1999, 14: 1371.   \href{https://arxiv.org/abs/gr-qc/9904026} {arXiv: gr-qc/9904026}

\bibitem{cha5}
S. Das, E. C. Vagenas,  Phys. Rev. Lett. 2008, 101: 221301. \href{http://dx.doi.org/10.1103/PhysRevLett.101.221301} {DOI: 10.1103/PhysRevLett.101.221301}

\bibitem{cha6}
S. Ghosh,  Class. Quantum Gravity 2014, 31:  025025. \href{https://arxiv.org/abs/1303.1256}{arXiv:1303.1256}

\bibitem{cha7}
F. Scardigli, R.  Casadio,  Eur. Phys. J. C 2015, 75: 425.  \href{https://arxiv.org/abs/1407.0113}{arXiv:1407.0113}

\bibitem{cha8}
D. Gao, M.  Zhan,  Phys. Rev. A 2016, 94: 013607. \href{https://arxiv.org/abs/1607.04353}{arXiv:1607.04353}

\bibitem{cha9}
Z.-W. Feng,  S.-Z Yang, H.-L. Li,  X.-T.  Zu, Phys. Lett. B 2017,  768: 81. \href{https://arxiv.org/abs/1610.08549}{arXiv:1610.08549}

\bibitem{cha10}
S. Kouwn, Phys. Dark Universe 2018, 21: 76. \href{https://arxiv.org/abs/1805.07278}{arXiv:1805.07278}

\bibitem{cha11}
P. Bushev, J. Bourhill, M. Goryachev, N. Kukharchyk, E. Ivanov, S.  Galliou, M.  Tobar, S. Danilishin, Phys. Rev. D 2019, 100: 066020.  \href{https://arxiv.org/abs/1903.03346}{arXiv: 1903.03346}

\bibitem{cha12}
S. Giardino, V. Salzano,  Eur. Phys. J. C 2021, 81: 110.  \href{https://arxiv.org/abs/2006.01580}{arXiv:2006.01580}

\bibitem{cha13}
J. C. S. Neves,  Eur. Phys. J. C 2020, 80, 343. \href{https://arxiv.org/abs/1906.11735}{arXiv:1906.11735}

\bibitem{cha13+}
\"{O}. \"{O}kc\"{u}, E. Aydiner,  Nucl. Phys. B 2021, 964, 115324. \href{http://dx.doi.org/10.1016/j.nuclphysb.2021.115324} {DOI: 10.1016/j.nuclphysb.2021.115324}

\bibitem{cha14}
F. Marin, F. Marino, M. Bonaldi, M. Cerdonio, L. Conti, P. Falferi, R. Mezzena, A. Ortolan, G. A. Prodi, L. Taf\/farello, G.  Vedovato, A. Vinante, J.-P. Zendri,  Nat. Phys. 2013, 9: 71. \href{http://dx.doi.org/10.1038/nphys2503} {DOI: 10.1038/nphys2503}

\bibitem{cha15}
M. Salah, F. Hammad, M. Faizal, A. F. Ali,   J. Cosmol. Astropart. Phys. 2017, 02: 035. \href{https://arxiv.org/abs/1608.00560}{arXiv:1608.00560}

\bibitem{cha16}
F. Scardigli,  G. Lambiase, E. C. Vagenas, Phys. Lett. B 2017, 767: 242.  \href{https://arxiv.org/abs/1611.01469}{arXiv:1611.01469}

\bibitem{cha17}
L. Buoninfante, G. G. Luciano, L.  Petruzziello,  Eur. Phys. J. C 2019, 79: 663. \href{https://arxiv.org/abs/1903.01382}{arXiv:1903.01382}

\bibitem{cha18}
H. Moradpour, A. H. Ziaie, S. Ghaffari, F. Feleppa,  Mon. Not.  R  Astron. Soc. 2019, 488:  L69. \href{https://arxiv.org/abs/1907.12940}  {arXiv:1907.12940}

\bibitem{cha19}
R. Casadio, F.  Scardigli,  Phys. Lett. B   2020 ,  807: 135558.  \href{http://dx.doi.org/10.1016/j.physletb.2020.135558} {DOI: 10.1016/j.physletb.2020.135558}

\bibitem{cha20}
P. Chen, Y. C. Ong, D.-H. Yeom,  Phys. Rep. 2015, 603: 1.  \href{https://arxiv.org/abs/1412.8366} {arXiv:1412.8366}

\bibitem{chb20}
I. Sakalli, A. \"{O}vg\"{u}n, K. Jusufi,  Astrophys. Space Sci.  2016, 361: 330.  \href{https://doi.org/10.1007/s10509-016-2922-x} {DOI:10.1007/s10509-016-2922-x}

\bibitem{cha21}
Z. W. Feng, H.-L. Li, X.-T. Zu, S.-Z. Yang,  Eur. Phys. J. C  2016, 76: 212.  \href{https://arxiv.org/abs/1604.04702} {arXiv:1604.04702}

\bibitem{chc1}
Scardigli, F.;  Blasone, M.;  Luciano, G.;  Casadio, R. Eur. Phys. J. C 2018, 78: 728. \href{https://arxiv.org/abs/1804.05282} {arXiv:1804.05282}

\bibitem{cha22}
H.-L. Li, Z.-W. Feng, S.-Z. Yang, X.-T. Zu,  Eur. Phys. J. C 2018, 78: 768. \href{http://dx.doi.org/10.1140/epjc/s10052-018-6252-8} {DOI: 10.1140/epjc/s10052-018-6252-8}

\bibitem{chb22}
E. C. Vagenas, S. M. Alsaleh, A. F.Ali,  Europhys. Lett. 2018, 120: 40001. \href{https://doi.org/10.1209/0295-5075/120/40001} {DOI: 10.1209/0295-5075/120/40001}

\bibitem{chb23}
S. Kanzi, I. Sakalli,  Nucl. Phys. B 2019,  946: 114703. \href{https://doi.org/10.1016/j.nuclphysb.2019.114703} {DOI: 10.1016/j.nuclphysb.2019.114703}

\bibitem{cha23}
S. Barman,  Eur. Phys. J. C 2020,  80: 50. \href{https://arxiv.org/abs/1907.09228}  {arXiv:1907.09228}

\bibitem{cha24}
H. Hassanabadi,  E. Maghsoodi, W. S. Chung,  Eur. Phys. J. C  2019, 79: 358. \href{http://dx.doi.org/10.1140/epjc/s10052-019-6871-8} {DOI: 10.1140/epjc/s10052-019-6871-8}

\bibitem{cha25}
A. Iorio, G. Lambiase, P. Pais, F. Scardigli,  Phys. Rev. D 2020, 101: 105002. \href{http://dx.doi.org/10.1103/PhysRevD.101.105002} {DOI: 10.1103/PhysRevD.101.105002}

\bibitem{chb25}
K. Blanchette, S. Das, S. Rastgoo,  J. High Energy Phys. 2021,  2021: 62. \href{https://doi.org/10.1007/JHEP09(2021)062} {DOI: 10.1007/JHEP09(2021)062}

\bibitem{chb26}
S. Kanzi, I. Sakalli,   Eur. Phys. J Plus  2022, 137: 14. \href{https://doi.org/10.1140/epjp/s13360-021-02245-7} {DOI: 10.1140/epjp/s13360-021-02245-7}

\bibitem{chb27}
I. Sakalli, S. Kanzi,  Ann. Phys. 2022, 439: 168803. \href{https://doi.org/10.1016/j.aop.2022.168803} {DOI: 10.1016/j.aop.2022.168803}

\bibitem{chb28}
Y. Chen, H.-L. Li,  \href{https://arxiv.org/abs/2001.11193}  {arXiv:2001.11193}

\bibitem{chb29}
M.-S. Ma, Y.-S.Liu,   \href{https://arxiv.org/abs/1810.06955}  {arXiv:1810.06955}

\bibitem{cha26}
T. Zhu, J.-R. Ren, M.-F. Li,  Phys. Lett. B 2009, 674: 204. \href{https://arxiv.org/abs/0811.0212}  {arXiv:0811.0212}

\bibitem{cha27}
W. Chemissany, S. Das, A. F. Ali, E. C. Vagenas,   J. Cosmol. Astropart. Phys. 2011, 12: 017. \href{https://arxiv.org/abs/1111.7288}  {arXiv:1111.7288}

\bibitem{cha28}
K. Zeynali, F. Darabi, H. Motavalli,  J. Cosmol. Astropart. Phys.  2012, 12: 033. \href{https://arxiv.org/abs/1206.0891}  {arXiv:1206.0891}

\bibitem{cha29}
K. Atazadeh, F. Darabi,  Phys. Dark Universe 2017, 16: 87. \href{https://arxiv.org/abs/1701.00060}  {arXiv:1701.00060}

\bibitem{cha30}
S. Das, R. B.  Mann,  Phys. Lett. B 2011, 704: 596. \href{https://arxiv.org/abs/1109.3258}  {arXiv:1109.3258}

\bibitem{cha31}
H. Verma, T. Mitra, B. P. Mandal,   Europhys. Lett. 2018, 123: 30009.  \href{https://arxiv.org/abs/1808.00766}  {arXiv:1808.00766}

\bibitem{cha32}
E. C. Vagenas, L. Alasfar, S. M. Alsaleh, A. F.  Ali,  Nucl. Phys.  B 2018, 931: 72. \href{https://arxiv.org/abs/1706.06502}  {arXiv:1706.06502}

\bibitem{cha33}
D. Park, E. Jung,   Phys. Rev. D 2020, 101: 066007.  \href{https://arxiv.org/abs/2001.02850}  {arXiv:2001.02850}

\bibitem{cha59}
P. Jizba, H.  Kleinert, F. Scardigli,  Phys. Rev. D 2010, 81: 084030. \href{https://arxiv.org/abs/0912.2253}  {arXiv:0912.2253}

\bibitem{cha34}
B. J. Carr, J.  Mureika, P. Nicolini,  J. High Energy Phys. 2015, 07: 052. \href{https://arxiv.org/abs/1504.07637}  {arXiv:1504.07637}

\bibitem{cha35+}
A. Chatterjee, A. Ghosh,  Phys. Rev. Lett. 2020, 125: 041302. \href{https://arxiv.org/abs/2007.15401}  {arXiv:2007.15401}

\bibitem{cha36}
M. Moussa,  Adv. High Energy Phys. 2015, 2015: 343284. \href{https://arxiv.org/abs/1512.04337}  {arXiv:1512.04337}

\bibitem{cha37}
R. Rashidi,  Ann. Phys. 2016, 374: 434. \href{https://arxiv.org/abs/1512.06356}  {arXiv:1512.06356}

\bibitem{cha35}
Y. C.  Ong, J. Cosmol. Astropart. Phys. 2018, 09: 015. \href{https://arxiv.org/abs/1804.05176}  {arXiv:1804.05176}

\bibitem{cha38}
Y. C.  Ong, J. High Energy Phys. 2018, 10: 195. \href{https://arxiv.org/abs/1806.03691}  {arXiv:1806.03691}

\bibitem{cha38+}
D. Kastor, S. Ray, J. Traschen,  Class. Quantum Gravity 2009, 26: 195011. \href{https://arxiv.org/abs/0904.2765} {arXiv:0904.2765}

\bibitem{cha38++}
D. Kastor, S. Ray, J. Traschen,  Class. Quantum Gravity 2011, 28:  195022. \href{https://arxiv.org/abs/1106.2764} {arXiv:1106.2764}

\bibitem{cha39}
S. W. Hawking, D. N. Page,  Commun. Math. Phys. 1983, 87: 577.  \href{http://dx.doi. org/10.1007/BF01208266} {DOI: 10.1007/BF01208266}

\bibitem{cht1}
S. A. Hartnoll,  Class. Quantum Gravity  2009, 26: 224002.  \href{https://arxiv.org/abs/0903.3246}  {arXiv:0903.3246}

\bibitem{cht2}
 J. McGreevy, Adv. High Energy Phys. 2010, 2010: 723105. \href{https://arxiv.org/abs/0909.0518}  {arXiv:0909.0518}

\bibitem{cha40}
D. Kubiz\v{n}\'{a}k, R. B. Mann,   J. High Energy Phys. 2012, 07: 033. \href{https://arxiv.org/abs/1205.0559}  {arXiv:1205.0559}

\bibitem{cha41}
S.-W. Wei, Y.-X. Liu,  Phys. Rev. Lett. 2015, 115: 111302.  \href{https://arxiv.org/abs/1502.00386}  {arXiv:1502.00386}

\bibitem{cha42}
S. H. Hendi, R. B. Mann, S. Panahiyan, B. E. Panah,  Phys. Rev. D 2017, 95: 021501.  \href{https://arxiv.org/abs/1702.00432}  {arXiv:1702.00432}

\bibitem{cha43}
M.-S. Ma, R.-H.  Wang,  Phys. Rev. D, 2017, 96: 024052.  \href{https://arxiv.org/abs/1707.09156}  {arXiv:1707.09156}

\bibitem{cha44}
A. Dehyadegari, A. Sheykhi, Phys. Rev. D 2017, 98: 024011.  \href{https://arxiv.org/abs/1711.01151}  {arXiv:1711.01151}

\bibitem{cha45}
S.-W. Wei, Y.-X. Liu,  Phys. Rev. D 2020, 101: 104018. \href{https://arxiv.org/abs/2003.14275}  {arXiv:2003.14275}

\bibitem{cha46}
M. Rostamia,  J. Sadeghibac, S.  Miraboutalebi,  Phys. Dark Universe 2020, 29: 100590.  \href{http://dx.doi.org/10.1016/j.dark.2020.100590} {DOI: 10.1016/j.dark.2020.100590}

\bibitem{cha46+}
M. Rostamia, J.  Sadeghibac, S. Miraboutalebi,  Phys. Rev. D 2020,  101: 044001.  \href{http://dx.doi.org/10.1103/PhysRevD.101.044001} {DOI: 10.1103/PhysRevD.101.044001}

\bibitem{chxx58}
 J. W. York,   Phys. Rev. D  1986,  33: 2092. \href{http://dx.doi.org/10.1103/PhysRevD.33.2092} {DOI: 10.1103/PhysRevD.33.2092}

\bibitem{chb49}
H. W. Braden, J. Brown, B. F. Whiting, J. W. York,  Phys. Rev. D  1990,  42: 3376.  \href{http://dx.doi.org/10.1103/PhysRevD.33.2092} {DOI: 10.1103/PhysRevD.42.3376}

\bibitem{chb52}
J. X. Lu, S. Roy, Z. Xiao,   J. High Energy Phys. 2011, 2011: 133. \href{https://arxiv.org/abs/1010.2068}  {arXiv:1010.2068}

\bibitem{chxx59}
J. X. Lu, R. Wei,  J. High Energy Phys. 2013 , 2013: 134. \href{https://arxiv.org/abs/1301.1780}  {arXiv:1301.1780}

\bibitem{chxx60}
D. Zhou, Z. Xiao,  J. High Energy Phys. 2015 , 2015: 134. \href{http://dx.doi.org/10.1088/0264-381/6/12/018} {DOI: 10.1088/0264-381/6/12/018}

\bibitem{chxx61}
Y.  Peng,  Phys. Lett. B 2018, 780: 144. \href{https://arxiv.org/abs/1801.02495}  {arXiv:1801.02495}

\bibitem{chxx62}
B. Kiczek, M. Rogatko,  Phys. Rev. D 2020, 101: 084035. \href{https://arxiv.org/abs/2004.06617}  {arXiv:2004.06617}

\bibitem{cha47}
D. Pav\'{o}n,  Phys. Rev. D  1991, 43: 2495. \href{http://dx.doi.org/10.1088/0264-381/6/12/018} {DOI: 10.1088/0264-381/6/12/018}

\bibitem{cha48}
P. C. W. Davies,  Class. Quantum Gravity   1989,  6: 1909. \href{http://dx.doi.org/10.1103/PhysRevD.43.2495} {DOI: 10.1103/PhysRevD.43.2495}

\bibitem{cha49}
 D. J. Gross, M. J.  Perry, L. G. Yaffe,  Phys. Rev. D 1982, 25: 330. \href{http://dx.doi.org/10.1103/PhysRevD.43.2495} {DOI: 10.1103/PhysRevD.43.2495}

\bibitem{chb47}
R. Mandal, S. Bhattacharyya, S. Gangopadhyay,  Gen. Relat. Gravit.  2018,  50: 143. \href{https://arxiv.org/abs/1805.07005 }  {arXiv:1805.07005 }

\bibitem{chb48}
S. Gangopadhyay, A. Dutta,   Adv. High Energy Phys.   2018,  2018: 7450607. \href{https://arxiv.org/abs/1805.11962}  {arXiv:1805.11962}

\bibitem{chb49+}
N. Kumar, S. Bhattacharyya. S. Gangopadhyay,  Gen. Relat. Gravit.  2020, 52: 20. \href{https://arxiv.org/abs/1904.13059}  {arXiv:1904.13059}

\bibitem{chb50+}
S. Haroon, R. A. Hennigar, R. B. Mann, F. Simovic,  Phys. Rev. D  2020,  101: 084051. \href{https://arxiv.org/abs/2002.01567}  {arXiv:2002.01567}

\bibitem{chb51+}
C. Promsiri, E. Hirunsirisawat, W. Liewrian,  Phys. Rev. D  2021, 104: 064004. \href{https://arxiv.org/abs/2106.02406 }  {arXiv:2106.02406 }

\bibitem{chb52+}
A. Dehyadegari, A. Sheykhi,  Phys. Rev. D  2021, 104: 104066. \href{https://arxiv.org/abs/2107.02915}  {arXiv:2107.02915}

\bibitem{chb53+}
Y. Liu, ;H.-D. Lyu, A. Raju  J. High Energy Phys.  2021,  2021: 140. \href{https://arxiv.org/abs/2108.04554}  {arXiv:2108.04554}

\bibitem{chb50}
S. Carlip, S. Vaidya,  Class. Quantum Gravity  2001,  20: 3827. \href{https://arxiv.org/abs/gr-qc/0306054}  {arXiv:gr-qc/0306054}

\bibitem{chb51}
A. P. Lundgren,  Phys. Rev. D 2008, 77: 044014. \href{https://arxiv.org/abs/gr-qc/0306054}  {arXiv:gr-qc/0612119}

\bibitem{chb53}
S. R. Dolan, S. Ponglertsakul, E. Winstanley,  Phys. Rev. D 2015,  92: 124047. \href{https://arxiv.org/abs/1507.02156}  {arXiv:1507.02156}

\bibitem{chb54}
S.  Ponglertsakul, E. Winstanley, S. R.  Dolan, Phys. Rev. D 2016, 94: 024031. \href{https://arxiv.org/abs/1604.01132}  {arXiv:1604.01132}

\bibitem{chb55}
N. Sanchis-Gual,  J. C Degollado, C. Herdeiro, J. A.  Font, P. J.  Montero,  Phys. Rev. D 2016,  94: 044061. \href{https://arxiv.org/abs/1607.06304}  {arXiv:1607.06304}

\bibitem{cha58+}
P. Wang, H.  Yang, S.  Ying,  Phys. Rev. D 2020,  101:  064045.  \href{https://arxiv.org/abs/1909.01275}  {arXiv:1909.01275}

\bibitem{cha58++}
P. Wang, H.  Wu, H.  Yang,  J. High Energy Phys. 2019, 07:  2.  \href{https://arxiv.org/abs/1901.06216}  {arXiv:1901.06216}

\bibitem{cha59++}
P. Wang, H. Wu, H. Yang, F. Yao,   J. High Energy Phys. 2020, 2019: 154.   \href{https://arxiv.org/abs/2006.14349}  {arXiv:2006.14349}

\bibitem{cha60++}
F. Simovic,  R. B.Mann,  Class. Quantum Gravity 2019, 36:  014002. \href{https://arxiv.org/abs/1807.11875}  {arXiv:1807.11875}

\bibitem{cha61++}
S. Haroon, R. A. Hennigar, R. B.  Mann, F.  Simovic, Phys. Rev. D 2020, 101: 084051. \href{https://arxiv.org/abs/2002.01567}  {arXiv:2002.01567}

\bibitem{chf1}
S.-W. Wei, Y.-X. Liu,  R. B. Mann, Phys. Rev. D 2019, 100: 124033. \href{https://doi.org/10.1088/1361-6382/aaf445} {DOI: 10.1088/1361-6382/aaf445}

\bibitem{chf2}
F. Simovic, R. B. Mann,  Class. Quantum Gravity 2019, 36: 014002. \href{https://doi.org/10.1088/1361-6382/aaf445} {DOI: 10.1088/1361-6382/aaf445}

\bibitem{cha49+}
 Y. Gim, W.  Kim,   J. Cosmol. Astropart. Phys. 2014, 10:  003.  \href{https://arxiv.org/abs/1406.6475}  {arXiv:1406.6475}

\bibitem{cha50}
Y. Gim, W. Kim,   J. Cosmol. Astropart. Phys. 2015, 05:  002. \href{https://arxiv.org/abs/1501.04702}  {arXiv:1501.04702}

\bibitem{cha51}
Z.-W. Feng, S.-Z. Yang, Phys. Lett. B 2017, 772:  737. \href{https://arxiv.org/abs/1708.06627}  {arXiv:1708.06627}

\bibitem{cha52}
Y.-W. Kim, S. K.  Kim, Y.-J. Park,  Eur. Phys. J. C 2016, 76:  557. \href{https://arxiv.org/abs/1607.06185}  {arXiv:1607.06185}

\bibitem{cha53}
S. Upadhyay, S. H.  Hendi, S. Panahiyan, B. E. Panah,  Prog.  Theor. Exp. Phys. 2018, 2018: 093E01. \href{https://arxiv.org/abs/1809.01078}  {arXiv:1809.01078}

\bibitem{cha54}
M. Shahjalal,  Phys. Lett. B 2018, 784: 6. \href{http://dx.doi.org/10.1016/j.physletb.2018.07.032} {DOI: 10.1016/j.physletb.2018.07.032}

\bibitem{cha55}
Z.-W. Feng, D.-L. Tang, D.-D. Feng, S.-Z. Yang,   Mod. Phys. Lett.  A 2020, 35:  2050010. \href{https://arxiv.org/abs/1710.04496}  {arXiv:1710.04496}

\bibitem{cha56}
M.-S. Ma,Y.-S.  Liu,  Adv. High Energy Phys. 2018, 2018: 1257631. \href{https://arxiv.org/abs/1810.06955}  {arXiv:1810.06955}

\bibitem{cha57}
Y. Chen, H.-L.  Li,  \href{https://arxiv.org/abs/2001.11193}  {arXiv:2001.11193}

\bibitem{cha58}
 P. Jizba, H. Kleinert. F. Scardigli,   Phys. Rev. D 2010, 81: 084030. \href{https://arxiv.org/abs/0912.2253}  {arXiv:0912.2253}

\bibitem{cha59x}
T. Kanazawa, G. Lambiase, G. Vilasi, A. Yoshioka,  Eur. Phys. J. C  2019, 79: 95. \href{http://dx.doi.org/10.1140/epjc/s10052-019-6610-1} {DOI:10.1140/epjc/s10052-019-6610-1}

\bibitem{cha60+}
V. Nenmeli, S. Shankaranarayanan, V. Todorinov, S. Das,  Phys. Lett. B 2021, 821: 136621. \href{https://arxiv.org/abs/2106.04141}  {arXiv:2106.04141}

\bibitem{cha61+}
S. Das, M. Fridman, G. Lambiase, E. C. Vagenas,  Phys. Lett. B 2021,  824: 136841. \href{https://arxiv.org/abs/2107.02077}  {arXiv:2107.02077}

\bibitem{chp4}
R. J.  Adler, P. Chen, D. I.  Santiago,  Gen. Rel. Grav. 2001, 33:  2101.  \href{https://arxiv.org/abs/gr-qc/0106080}  {arXiv: gr-qc/0106080}

\bibitem{chp5}
D. Y. Chen, H. W. Wu, H. T. Yang,  Adv. High Energy Phys. 2013, 2013: 432412.   \href{https://arxiv.org/abs/1305.7104}  {arXiv:1305.7104}

\bibitem{chp6}
D. Y. Chen, Q. Q.  Jiang, P.  Wang, H. Yang,   J. High Energy Phys. 2012, 2013:  176.  \href{http://dx.doi.org/10.1007/JHEP11(2013)176} {DOI:10.1007/JHEP11(2013)176}

\bibitem{chp7}
Z. Feng, L. Zhang, X.  Zu,  Mod. Phys. Lett. A  2014, 29: 1450123. \href{http://dx.doi.org/10.1142/S0217732314501235} {DOI:10.1142/S0217732314501235}

\bibitem{chp8}
S. Gunasekaran, R. B. Mann, D. Kubiz\v{n}\'{a}k,   J. High Energy Phys. 2012,   2012: 110.  \href{https://arxiv.org/abs/1208.6251}  {arXiv:1208.6251}

\bibitem{cha60}
Y.-W. Kim, S. K.  Kim, Y.-J.  Park,  Eur. Phys. J. C 2016, 76: 557. \href{http://dx.doi.org/10.1140/epjc/s10052-016-4393-1} {DOI: 10.1140/epjc/s10052-016-4393-1}

\bibitem{chb1}
R. Casadio, A. Orlandi,   J. High Energy Phys. 2013, 1308: 025. \href{https://arxiv.org/abs/1302.7138}  {arXiv:1302.7138}

\bibitem{chb2}
R. Casadio, A.  Giugno, O.  Micu, A. Orlandi,  Phys. Rev. D 2014,  90: 084040. \href{https://arxiv.org/abs/1405.4192}  {arXiv:1405.4192}

 \bibitem{chd1}
K. Nozari, S. H. Mehdipour, Mod. Phys. Lett. A 2005, 20: 2937. \href{https://arxiv.org/abs/0809.3144}  {arXiv:0809.3144}
\end{thebibliography}
\end{document}